\documentclass[reprint,nofootinbib,amsmath,amssymb,aps]{revtex4-1}
\usepackage{graphicx}
\usepackage{dcolumn}
\usepackage[colorlinks,linkcolor=magenta,anchorcolor=cyan,citecolor=blue]{hyperref}
\usepackage{bm}
\usepackage[multiple]{footmisc}

\def\be{\begin{equation}}
\def\ee{\end{equation}}
\def\ba{\begin{eqnarray}}
\def\ea{\end{eqnarray}}

\def\lf{\left}
\def\rt{\right}

\newcommand{\eq}[1]{(\ref{#1})}

\def\lf{\left}\def\rt{\right} \def\w{\omega}  \def\y {\psi}   \def\p {\pi}   \def\d {\delta} \def\f {\phi}    \def\k {\kappa} \def\l {\lambda} \def\z {\zeta} \def\x {\xi}  \def\b {\beta}   \def\pd {\partial}\def\p {\pi} \def \inf {\infty}  \def \e { \varepsilon}
\def\Q{\Theta}      \def\S {\Sigma}  \def\F {\Phi}      \def\grad{\nabla}\def\.{\cdot}
\def\math {\mathcal}
\begin{document}

\title{Weak cosmic censorship conjecture in higher-dimensional black holes with nonlinear electrodynamic sources }
\author{Zhen Li}
\email{zhen\_li@mail.bnu.edu.cn}
\affiliation {Department of Physics, Beijing Normal University, Beijing, 100875, China}
\author{Yunjiao Gao}
\email{201821140015@mail.bnu.edu.cn}
\affiliation {Department of Physics, Beijing Normal University, Beijing, 100875, China}
\author{Xiao-Kan Guo}
\email{Corresponding author. xkguo@bnu.edu.cn}
\affiliation {Department of Physics, Beijing Normal University, Beijing, 100875, China}
\date{\today}

\begin{abstract}
The new version of the gedanken experiments proposed by Sorce and Wald are designed to test the validity of the weak cosmic censorship conjecture (WCCC) by overspinning or overcharging the Kerr-Newman black hole in Einstein-Maxwell gravity. Following their setup, in this paper, we investigate the WCCC in the higher-dimensional charged black hole with a nonlinear electrodynamics source. We derive the first and seconder order perturbation inequalities of the charged collision matter based on the Iyer-Wald formalism as well as the null energy conditions, and show that they share the similar form as that in Einstein-Maxwell gravity. As a result, we find that the static higher-dimensional nonlinear electrodynamics (HDNL) black holes cannot be overcharged after considering these two inequalities. Our result might indicate the validity of WCCC for more general HNDL and related systems.
\end{abstract}
\maketitle

\section{Introduction}\label{sec1}

In general relativity, there are gravitational singularities inside black holes. These singularities are not problematic as long as they are hided by the even horizons.
The naked singularity, however,  will invalidate causality of spacetime. In order to  eliminate the possibility of a naked singularity, Penrose formulated the weak cosmic censorship conjecture (WCCC) that any physical process cannot destroy the event horizon of the black holes \cite{RPenrose}.

There have been many efforts made over the years to either prove or disprove WCCC. A famous work is the gedanken experiment proposed by Wald \cite{Wald97,Wald94}. In \cite{Wald97,Wald94} he considers the overspinning or overcharging of an extremal Kerr-Newman (KN) black hole by throwing in a test particle and find that an extremal Kerr-Newman black hole cannot be destroy in such a way. However, Hubeny \cite{Hubeny} reaches the opposite conclusion by considering a special particle with charge. This contradiction has drawn a lot of attentions \cite{1,2,3,4,5,B1,B2,B3,B4,B5,B6,B7,B8,B9,B10,B11,B12,B13,B14,B15,B16,B17,SG,RV}.

In the above gedanken experiments, the self-force and finite-size effects for a given test particle are ignored, and the corrections from the conserved charges are only up to the first-order perturbation. To solve these problems, Sorce and Wald have recently proposed a new version of gedanken experiment without relying on the test particle assumption \cite{SW}. In this new gedanken experiment, the second-order corrections from the mass, angular momentum and charge are taken  into account, and the first two perturbation inequalities of the collision matters are also obtained in the Iyer-Wald formalism \cite{IW}. Given these, they show that WCCC is still valid for the KN black holes.

Recently, this new version of gedanken experiment has also been generalized to many other cases \cite{FMP,HDR,CD,KS,WJ,EBI,GB,EM,RBH,HRN,LL,DL,HDADS,S1,S2,S3,RG}, and WCCC is shown to be valid for these black holes. Since a general proof of WCCC is still lacking, it is necessary to investigate the gedanken experiments in various theories, especially for those with particular characteristics or more general properties. The option of considering higher-dimensional charged black holes with a nonlinear electrodynamics source (HDNL black holes) is motivated by the fact that nonlinear electrodynamics models provide us with an excellent laboratory for constructing black hole solutions with interesting properties \cite{HA}, for instance, regular black holes \cite{RRG,R1,R2,R3,R4,R5,R6,R7,R8}. Although the HDNL black holes are possibly regular, the naked singularities are not prohibited in all cases \cite{N1,N2,N3,N4,N5,N6}. In this paper, we consider the HDNL black holes with gravitational singularities, and for such HDNL black holes we study the new gedanken experiment as well as the WCCC.\\
\indent This paper is organized as follows. In section.\ref{sec2}, we briefly review the Iyer-Wald formalism for general diffeomorphism-covariant theories and give the corresponding variational quantities. In section.\ref{sec3}, we restrict ourselves on the HDNL theory, where the explicit expressions for the corresponding conserved charges are presented and the related quantities of the static HDNL black hole are introduced. In section.\ref{sec4}, we present the setup for the new version of gedanken experiment for the HDNL black holes, and derive the first and second order inequalities for the perturbations. In section.\ref{sec5}, we conduct the new version of gedanken experiment to the nearly extremal HNDL black holes, and verify that no violation of WCCC can occur when the second-order perturbation inequality is considered. Finally, the conclusions are presented in section.\ref{sec6}.

\section{Iyer-Wald formalism in a diffeomorphism-covariant gravity}\label{sec2}
 Recently, Sorce and Wald proposed a new version of the gedanken experiments by using the Iyer-Wald formalisms \cite{IW}. This method is based on the variational theory of an action on a general diffeomorphism-covariant n-dimensional oriented manifold $\bm{\math{M}}$. The Lagrangian $\bm{L}$ is a $n$-form field that depends on locally the metric $g_{ab}$ and other fields $\y$. Following the notation in \cite{SW}, we use $\f=(g_{ab},\y)$ to denote all dynamical fields. By performing a variation to the Lagrangian $\bm{L}$, we have
\ba\begin{aligned}\label{varL}
\d \bm{L}=\bm{E}_\f\d\f+d\bm{\Q}(\f,\d\f).
\end{aligned}\ea
The equations of motion (EoM) will be given by $\bm{E}_\f=0$, and $\bm{\Q}$ is the symplectic potential three-form. The symplectic current three-form is
\ba\begin{aligned}
\bm{\w}(\f,\d_1\f, \d_2\f)=\d_1\bm{\Q}(\f,\d_2\f)-\d_2\bm{\Q}(\f,\d_1\f).
\end{aligned}\ea
The Noether current three-form $\bm{J}_\z$ associated with a vector field $\z^a$ is then
\ba\label{defJ}
\bm{J}_\z=\bm{\Q}(\f, \math{L}_\z\f)-\z\.\bm{L}\equiv\bm{C}_\z+d\bm{Q}_\z
\ea
where $\bm{Q}_\z$ is the  Noether charge two-form related to $\z^a$ and $\bm{C}_\z=\z^a\bm{C}_a$ are the constraints of the theory, i.e., the on-shell dynamical fields will give us $\bm{C}_a=0$.

With the help of  diffeomprphism invariance, $\z^a$ can be fixed. If $\z^a$ is a Killing vector field and  the background fields satisfy the EOM, we can further obtain the first two variation identities,
\begin{align}
d[\d\bm{Q}_\z-\z\.\bm{\Q}(\f,\d\f)]=&\bm{\w}\lf(\f,\d\f,\math{L}_\z\f\rt)-\z\.\bm{E}\d\f-\d \bm{C}_\z,\label{var1}\\
d[\d^2\bm{Q}_\z-\z\.\d\bm{\Q}(\f,\d\f)]=&\bm{\w}\lf(\f,\d\f,\math{L}_\z\d\f\rt)-\nonumber\\
&-\z\.\d\bm{E}\d\f-\d^2 \bm{C}_\z.\label{var1b}
\end{align}

We are interested in the static HDNL black holes. Therefore, we assume that the background spacetime is asymptotic flat and static, and there exist a timelike Killing vector field $\x^a$ which is normalized at asymptotic infinity. By utilizing this Killing vector, the ADM mass of this black hole can be expressed as
\ba\begin{aligned}
\d M&=\int_\inf \lf[\d\bm{Q}_\x-\x\.\bm{\Q}(\f,\d\f)\rt].
\end{aligned}\ea
By replacing $\z^a$ with $\x^a$ and integrating the perturbation identities \eq{var1}\eq{var1b} on a hypersurface $\S$ with a cross section $\mathcal{B}$ of the horizon, we have
\begin{align}
\d M=&\int_{\mathcal{B}}\lf[\d\bm{Q}_\x-\x\.\bm{Q}(\f,\d\f)\rt]-\int_\S \d \bm{C}_\x,\\
\d^2 M=&\int_{\mathcal{B}}\lf[\d^2\bm{Q}_\x-\x\.\d \bm{Q}(\f,\d\f)\rt]-\nonumber\\
&-\int_\S\x\.\d \bm{E}\d\f-\int_\S \d \bm{C}_\x+\math{E}_\S(\f,\d\f),\label{var12}
\end{align}
where
\ba
\math{E}_\S(\f,\d\f)=\int_\S\bm{\w}(\f,\d\f,\math{L}_\x\d\f).
\ea

\section{HDNL gravity and its static solution}\label{sec3}
In this section, we consider the $n$-dimensional NL theory \cite{HA}  and put it into the Iyer-Wald form. The Lagrangian is
\ba
\bm{L}=\frac{1}{16\p}\bm{\epsilon}R-\alpha \bm{\epsilon}{\left( {{F_{ab}}{F^{ab}}} \right)^q}
\ea
where $\bm{\epsilon}\equiv {\epsilon_{{a_1} \cdot  \cdot  \cdot {a_n}}}$ is the volume element, ${F_{ab}} = {\nabla _a}{A_b} - {\nabla _b}{A_a}$ (or $\bm{F} = d\bm{A}$) is the field strength (Maxwell tensor). $\alpha $ is the coupling constant whose sign must be chosen such that the energy density is positive. This condition selects two branches depending on the range of the exponent $q$\cite{HA},
\begin{align}
{\mathop{\rm sgn}} (\alpha ) &=  - {( - 1)^q}\qquad\, for \quad q > \frac{1}{2}\nonumber\\
{\mathop{\rm sgn}} (\alpha ) &= {( - 1)^q}\qquad\quad for \quad q < \frac{1}{2}
\end{align}
Also, the exponent $q$ will be restricted in the following.

Accordingly, the symplectic potential is
\ba\begin{aligned}
\bm{\Q}(\f,\d\f)=\bm{\Q}^\text{GR}(\f,\d\f)+\bm{\Q}^{\text{NL}}(\f,\d\f)
\end{aligned}\ea
with components
\begin{align}
{\Q}_{{a_2} \cdot  \cdot  \cdot {a_n}}^\text{GR}(\f,\d\f)&=\frac{1}{16\p}\epsilon_{d{a_2} \cdot  \cdot  \cdot {a_n}}g^{de}g^{fg}\lf(\grad_g \d g_{ef}-\grad_e\d g_{fg}\rt)\label{qqq}\\
{\Q}_{{a_2} \cdot  \cdot  \cdot {a_n}}^\text{NL}(\f,\d\f)&=-\frac{1}{4\p}\epsilon_{d{a_2} \cdot  \cdot  \cdot {a_n}}B^{de}\delta A_e.\label{qqqq}
\end{align}
Here, we have denoted
$
{B^{de}} \equiv 16\pi \alpha q{F^{q - 1}}{F^{de}}$ and $ F \equiv {F_{ab}}{F^{ab}}
$.
The Noether charge is given by
\ba
\bm{Q}_\x=\bm{Q}_\x^\text{GR}+\bm{Q}_\x^\text{NL}\,
\ea
whose the components are
\begin{align}
\lf({Q}_\x^\text{GR}\rt)_{{a_3} \cdot  \cdot  \cdot {a_n}}&=-\frac{1}{16\p}\epsilon_{de{a_3} \cdot  \cdot  \cdot {a_n}}\grad^d\x^e,\\
\lf({Q}_\x^\text{NL}\rt)_{{a_3} \cdot  \cdot  \cdot {a_n}}&=-\frac{1}{8\p}\epsilon_{de{a_3} \cdot  \cdot  \cdot {a_n}}B^{de}\delta A_f\x^f.
\end{align}
The constraints can be shown to be
\ba\begin{aligned}\label{EC}
{C}_{f{a_2} \cdot  \cdot  \cdot {a_n}}&=\epsilon_{e{a_2} \cdot  \cdot  \cdot {a_n}}\lf(T_f{}^e+A_f j ^e\rt).
\end{aligned}\ea
Here we denote
\ba\begin{aligned}\label{TJ}
T_{ab}=\frac{1}{8\p}G_{ab}-T_{ab}^\text{NL},\quad
j^a=\frac{1}{4\p}\grad_b B^{ab}\,
\end{aligned}\ea
where $G_{ab}$ is the Einstein tensor, and the electromagnetic stress-energy tensor is
\ba\begin{aligned}\label{TTT}
T_{ab}^\text{NL}&= 4\alpha [q{F_{ac}}F_b^c{F^{q - 1}} - \frac{1}{4}{g_{ab}}{F^q}].
\end{aligned}\ea
 $T^{ab}$ and $j^a$ are nonvanishing when there are other charged matter sources;
 they represent the stress-energy tensor and the Maxwell charge current of the additional matter. Constraining $T^{ab}=j^a=0$ gives the EoM of the on-shell fields.

 We only consider the situation that background spacetime is static, that to say, $\math{L}_\x \bm{A}=\x\. \bm{F}+d\lf(\x\.\bm{A}\rt)=0$. Since $\x\. \bm{A}$ is a constant on the horizon, we have
\ba\label{Fform}
{\x_a}{B^{ab}}\propto\x_{a}F^{ab}\propto \x^b
\ea
on the horizon. From \eq{qqq} and \eq{qqqq}, the symplectic current for the HDNL theory can be written as
\ba
\bm{\w}(\f, \d_1\f,\d_2\f)=\bm{\w}^\text{GR}+\bm{\w}^\text{NL},
\ea
where, in components,
\begin{align}\label{3w}
{\w}_{{a_2} \cdot  \cdot  \cdot {a_n}}^\text{GR}=&\frac{1}{16\p}\epsilon_{d{a_2} \cdot  \cdot  \cdot {a_n}}w^d,\\
{\w}_{{a_2} \cdot  \cdot  \cdot {a_n}}^\text{NL}=&\frac{1}{4\p}\lf[\d_2\lf(\epsilon_{d{a_2} \cdot  \cdot  \cdot {a_n}}B^{de}\rt)\d_1 A_e- \right.\nonumber\\
&\qquad\,\,\left.-\d_1\lf(\epsilon_{d{a_2} \cdot  \cdot  \cdot {a_n}}B^{de}\rt)\d_2 A_e\rt],
\end{align}
with
\begin{align*}
&w^a=(g^{ae}g^{fb}g^{cd}-\frac{1}{2}g^{ad}g^{be}g^{fc}-\frac{1}{2}g^{ab}g^{cd}g^{ef}-\nonumber\\
-&\frac{1}{2}g^{bc}g^{ae}g^{fd}+\frac{1}{2}g^{bc}g^{ad}g^{ef})\lf(\d_2g_{bc}\grad_d\d_1 g_{ef}-\d_1 g_{bc}\grad_d\d_2g_{ef}\rt)
\end{align*}

Next, We focus on the static HDNL black hole solution in this theory. In addition, we only consider a purely radial electric Ansatz for the electromagnetic field which means that the only non-vanishing component of the Maxwell tensor is given by $F_{tr}$ \cite{HA}. Then, the line element and  the non-vanishing Maxwell tensor component are respectively \cite{HA}
\begin{align}
ds^2&=-f(r)dt^2+\frac{1}{f(r)}dr^2+{r^2}d\Omega _{n - 2}^2,\label{ds2}\\
F_{tr} &= \frac{C}{{{r^{\frac{{n - 2}}{{2q - 1}}}}}}\label{ele},
\end{align}
where $d\Omega _{n - 2}^2$ is the metric of a $(n-2)$-dimensional unit sphere ${S^{D - 2}}$, and
\ba\begin{aligned}
f(r) = 1 - \frac{m}{{{r^{n - 3}}}} + \frac{D}{{{r^\beta }}}.
\end{aligned}\ea
Here we have defined
\begin{equation}
D = \frac{{\kappa \alpha {{( - 2)}^{q + 1}}{{(2{\rm{q}} - 1)}^2}{C^{2q}}}}{{(n - 2)(n - 2{\rm{q}} - 1)}},\quad \beta  = \frac{{2(qn - 4q + 1)}}{{2q - 1}},
\end{equation}
 with $\kappa=8\p$. The $m$ and $C$ are two integration constants proportional to the mass and the electric charge of the black hole; their relations to the ADM mass $M$ and electric charge $Q$ of the black hole are as follow
\begin{align}\label{MQ}
m &= \frac{{16\pi M}}{{(n - 2)\Omega }}\\
{C^{2q - 1}} &= \frac{Q}{{4\alpha q{{( - 2)}^{q - 1}}\Omega }}\label{MQ2}
\end{align}
with $\Omega  = 2{\pi ^{(n - 1)/2}}/\Gamma [(n - 1)/2]$ being the volume of the unit $(n-2)$-sphere.

Since the only non-vanishing component of the Maxwell tensor is given by $F_{tr}$, the Maxwell invariant $F=-2(F_{tr})^2$ is negative, and the exponent $q$ can only be an integer or a rational number with odd denominator. Thus, in order to deal with real solutions, the exponent $q$ is restricted to be an element of the following set
\begin{equation}
\mathbb{Q} = \left\{ {\frac{m}{{2n + 1}}|\left( {m,n} \right) \in \mathbb{Z} \times \mathbb{Z}} \right\}.
\end{equation}

This metric solution suggests two natural ranges concerning the exponent $\beta$, namely whether  $\frac{1}{{{r^\beta }}}$ goes faster or not than the Schwarzschild potential $\frac{1}{{{r^{n - 3}}}}$ when r go to zero. When $0 < \beta  < n - 3$, the WCCC will certainly satisfied since $f(r)$ will change sign when $r$ close to zero. In other words, there will be a horizon. The cases $\beta  \le 0$ and $\beta = n - 3$ correspond to the non-asymptotically flat metric and the critical exponent $q = (n-1)/2$ respectively, and we will not consider these two cases in this paper. So, we only investigate the case when $\beta  > n - 3$. The condition $\beta > n - 3$ imposes the exponent $q$ to be in the following range $q \in \mathbb{Q} \cap (1/2,(n - 1)/2)$ which in turn implies that the  field strength (\ref{ele}) vanishes at infinity \cite{HA}. In order to have real roots for the metric function $f(r)$, the constant $m$ must be positive and the constant $D$ must be chosen in the following range
\begin{equation}
0 < D < (n - 3){\left( {\frac{m}{\beta }} \right)^{\frac{\beta }{{n - 3}}}}{(\beta  + 3 - n)^{\frac{{\beta  + 3 - n}}{{n - 3}}}}.
\end{equation}
Under these conditions, we have two roots localized at ${r_ - } \in (0,d)$ and ${r_ + } = {r_h} \in (d,\infty )$ where
\[
d = {\left( {\frac{m}{\beta }(\beta  + 3 - n)} \right)^{\frac{1}{{n - 3}}}}.
\]
The radius $r_h$ of the event horizon is the largest root of $f(r)=0$. The corresponding surface gravity, area of the black hole and electric potential are respectively
\begin{equation}\label{kAWF}
\k_h=\frac{f'(r_h)}{2}\,,\ A_\math{H}=\Omega r_h^{n-2},~
{\Phi _\math{H}}= \frac{1}{{3 - n + \beta }}\frac{C}{{r_h^{3 - n + \beta }}}.
\end{equation}
An extremal black hole can also be obtained if $m$ is positive and the constant $D$ given by
\ba\begin{aligned}\label{excond}
D = (n - 3){\left( {\frac{m}{\beta }} \right)^{\frac{\beta }{{n - 3}}}}{(\beta  + 3 - n)^{\frac{{\beta  + 3 - n}}{{n - 3}}}}
\end{aligned}\ea
with the horizon radius $r_e=d$ for the extremal HDNL black hole.

\section{Perturbation inequalities in HDNL gravity}\label{sec4}
Following  the new version of gedanken experiments proposed by Sorce and Wald \cite{SW}, in this section, we  derive the first and second order inequalities in the HDNL gravity. These two inequalities are essential for us to investigate whether WCCC is valid or not by virtue of the null energy condition of the matter sources.

We consider a one-parameter family additional charged matter source which is restricted to a compact region of the future horizon, and it will cause a perturbation to the static HDNL black hole. The EoM of the dynamical fields then can be written as
\begin{align}
R_{ab}(\l)-\frac{1}{2}R(\l)g_{ab}(\l)=&8\p \lf[T_{ab}^\text{NL}(\l)+T_{ab}(\l)\rt],\\
\nabla _b^{(\lambda )}B^{ab}(\lambda ) = &4\pi j^a.
\end{align}
We have set $T^{ab}=j^a=0$ for the background fields. Similar to \cite{SW}, we also assume the nearly extremal HDNL black holes with a bifurcate horizon to be linearly stable under perturbations. Therefore, we can always choose the hypersurface as $\S=\math{H}\cup \S_1$, and $\math{H}$ is a portion of the future horizon bounded by the bifurcate surface $\mathcal{B}$ as well as the late cross section $\mathcal{B}_1$ where the matter source vanishes. From $\mathcal{B}_1$ to the spatial infinity, there is a spacelike hypersurface $\S_1$ where the dynamical fields is described by the static HDNL solutions \eq{ds2}.

We know that the perturbation vanishes on the bifurcation surface $\mathcal{B}$. On the other hand, the background fields are source-free and satisfy the EOM $\bm{E}_\f=0$, then the first-order perturbation identity \eq{var12} reduce to
\ba\begin{aligned}\label{fst1}
\d M&=-\int_\math{H}\epsilon_{e{a_2} \cdot  \cdot  \cdot {a_n}}\lf[\delta T_f{}^e+A_f\delta j^e\rt]\x^f\,
\end{aligned}\ea
where we have used the fact that $T^{ab}=j^a=0$ for the background spacetime. As has been mentioned in section.\ref{sec3},  the electric field potential $\Phi = - {\xi ^a}{A_a}$ vanishes at asymptotic infinity, and it is a constant on the $\math{H}$. According to the HDNL part of \eq{TJ}, we  further obtain
\begin{align}
&-\int_\math{H}\epsilon_{e{a_2} \cdot  \cdot  \cdot {a_n}}\x^fA_f\d j^e=\F_\math{H}\d\lf[\int_\math{H}\epsilon_{e{a_2} \cdot  \cdot  \cdot {a_n}} j^e\rt]\nonumber\\
=&\frac{1}{4\p}\F_\math{H}\d\lf[\int_\math{H}\epsilon_{e{a_2} \cdot  \cdot  \cdot {a_n}} \grad_f B^{ef}\rt]\nonumber\\
=&\frac{1}{8\p}\F_\math{H}\d\lf[\int_{\mathcal{B}_1}\epsilon_{ed{a_3} \cdot  \cdot  \cdot {a_n}} B^{ed}-\int_{\mathcal{B}}\epsilon_{ed{a_3} \cdot  \cdot  \cdot {a_n}} B^{ed}\rt]\nonumber\\
=&\F_\math{H}\d Q
\end{align}
with the electric charge of the black hole given by
\ba
Q=\frac{1}{8\p}\int_\inf\epsilon_{ed{a_3} \cdot  \cdot  \cdot {a_n}} B^{ed}=\frac{1}{8\p}\int_{\mathcal{B}_1}\epsilon_{ed{a_3} \cdot  \cdot  \cdot {a_n}} B^{ed}.
\ea
Here we have used the Gaussian theorem and the current $j^a$ vanishes on $\S_1$. Using this result, Eq. \eq{fst1} becomes
\ba\label{var1eq1}\begin{aligned}
\d M-\F_\math{H}\d Q=\int_\math{H}\bm{\tilde{\epsilon}} \d T_{ab}k^a \x^b\geq 0
\end{aligned}\ea
where we have used the null energy condition $\d T_{ab}k^a k^b\geq 0$, and $\bm{\tilde{\epsilon}}$ is the volume element on the horizon obtained from $\epsilon_{e{a_2} \cdot  \cdot  \cdot {a_n}}=-4k_{[e}\tilde{\epsilon}_{{a_2} \cdot  \cdot  \cdot {a_n}]}$ with the future-directed normal vector on the horizon $k^a \propto \x^a$. According to (\ref{MQ}), (\ref{MQ2}) and (\ref{kAWF}), we also have
\begin{align}
\delta M - {\Phi _\math{H}}\delta Q = \frac{{(n - 2)\Omega }}{{16\pi }}\left( {\delta m - \frac{{\delta D}}{{r_h^{3 - n + \beta }}}} \right) \ge 0.
\end{align}

For the double-horizon black holes, if the naked singularity can be obtained by adding charges, it means that the optimal choice will make the HDNL black hole to saturate \eq{var1eq1}, i.e. $\d T_{ab}$ vanishes on the horizon. Then, we get
\ba\label{var1eq1op}\begin{aligned}
\d M-\F_\math{H}\d Q=0.
\end{aligned}\ea
Accordingly
\begin{equation}\label{ie11}
 {\delta m - \frac{{\delta D}}{{r_h^{3 - n + \beta }}}} = 0.
\end{equation}

Next, we consider the second-order perturbation inequality under this optimal condition. Similar  to the first-order analysis, we can  obtain
\begin{align}\label{sec22}
\d^2 M&=-\int_\math{H}\x\.\d \bm{E}_\f\d\f-\int_\math{H} \d \bm{C}_\x+\math{E}_\S(\f,\d\f)\nonumber\\
&=-\int_\math{H} \d \bm{C}_\x+\math{E}_\S(\f,\d\f)\,
\end{align}
where in last step, we have used the fact that $\x^a$ is tangent to the horizon. The integrals only depend on the $\math{H}$ since $\bm{E}_\f(\l)=\bm{C}(\l)=0$ on $\S_1$ (since the dynamical fields satisfy the source-free EOM on the hypersurface $\S_1$). With the optimal condition of the first-order perturbation as well as the energy condition for the second-order perturbed stress-energy tensor, Eq. \eq{sec22} reduces to
\begin{align}\label{dM22}
\d^2M-\F_\math{H}\d^2Q&=\math{E}_\S(\f,\d \f)+\int_\math{H}\bm{\tilde{\epsilon}}\d^2T_{ab}\x^ak^b\nonumber\\
&\geq\math{E}_\math{H}(\f,\d \f)+\math{E}_{\S_1}(\f,\d \f)\,
\end{align}
where we have imposed the condition $\x^a \d A_a|_\math{H}=0$ by a gauge transformation \cite{SW}. The first term of the right side in \eq{dM22} can be decomposed into
\ba\label{EEE3}
\math{E}_\math{H}(\f,\d \f)=\int_\math{H}\bm{\w}^\text{GR}+\int_\math{H}\bm{\w}^\text{NL}.
\ea
According to \cite{SW}, the gravitational part in above expression is given by
\ba
\int_\math{H}\bm{\w}^\text{GR}=\frac{1}{4\p}\int_{\math{H}}(\x^a\grad_a u)\d\rho_{ac}\d\rho^{bc}\bm{\tilde{\epsilon}}\geq 0.
\ea
where $\delta {\rho _{ab}}$ denotes the perturbed shear of the horizon generators, $u$ represents an affine parameter along the future horizon. For the HDNL part, according \eq{3w}, we have
\begin{align}\label{EEM}
&{\w}^\text{NL}_{{a_2} \cdot  \cdot  \cdot {a_n}}=\frac{1}{4\p}\epsilon_{d{a_2} \cdot  \cdot  \cdot {a_n}}\lf[\d A_e\math{L}_\x\d B^{de}-\d B^{de}\math{L}_\x\d A_e\rt]+\nonumber\\
&\qquad+\frac{1}{4\p}\lf[(\math{L}_\x\d\epsilon_{d{a_2} \cdot  \cdot  \cdot {a_n}})B^{de}\d A_e-\d\epsilon_{d{a_2} \cdot  \cdot  \cdot {a_n}}B^{de}\math{L}_\x\d A_e\rt].
\end{align}
By virtue of the gauge condition $\x^a\d A_a=0$ on the horizon as well as the assumption \eq{Fform}, the last two terms vanish. Then, Eq. \eq{EEM} becomes
\begin{align}\label{EMw}
{\w}^\text{NL}_{{a_2} \cdot  \cdot  \cdot {a_n}}=&\frac{1}{4\p}\math{L}_\x\lf(\epsilon_{d{a_2} \cdot  \cdot  \cdot {a_n}}\d  A_e\d B^{de}\rt)\nonumber\\
&-\frac{1}{2\p}\epsilon_{d{a_2} \cdot  \cdot  \cdot {a_n}}\d B^{de}\math{L}_\x\d A_e
\end{align}
By using the Stoke's theorem, the integral over $\math{H}$ of the first term will only contribute a boundary term at $\mathcal{B}_1$. According to the stability assumption, $\d B^{de}$ will also satisfies Eq.\eq{Fform}. With the help of the gauge condition $\x^a\d A_a=0$ on $\math{H}$, the first term of \eq{EMw} makes no contributions. Thus, we have
\begin{align}\label{EMw1}
\math{E}_\math{H}(\f,\d\f)&=-\frac{1}{2\p}\int_\math{H}\epsilon_{d{a_2} \cdot  \cdot  \cdot {a_n}}\d B^{de}\math{L}_\x\d A_e\nonumber\\
&=\int_\math{H}\bm{\tilde{\epsilon}}\x^ak^b\lf(\d^2T_{ab}^\text{NL}\rt)\geq 0
\end{align}
where we have used the null energy condition for the electromagnetic stress-energy tensor. Finally, \eq{dM22} reduces to
\ba
\d^2M-\F_\math{H}\d^2Q\geq \math{E}_{\S_1}(\f,\d\f).
\ea

Following \cite{SW}, we now  evaluate the  remaining term $\math{E}_{\S_1}(\f,\d\f)$. We first write $\math{E}_{\S_1}(\f,\d\f)=\math{E}_{\S_1}(\f,\d\f^\text{NL})$, where $\f^\text{NL}$ is introduced by the variation of a family of HDNL black hole solutions \eq{ds2},
\label{varMQ}\begin{align}
M^\text{NL}(\l)&=M+\l \d M\,\\
Q^\text{NL}(\l)&=Q+\l \d Q\,
\end{align}
where $\d M$ and $\d Q$ satisfy the first order optimal perturbation of the matter source. From the variation \eq{varMQ}, one can find $\d^2 M=\d^2 Q=\d \bm{E}=\d^2 \bm{C}=\math{E}_\math{H}(\f,\d \f^\text{NL})=0$. Thus, from \eq{var12}, we have
\ba
\math{E}_{\S_1}(\f, \d \f^\text{NL})=-\int_B\lf[\d^2\bm{Q}_\x-\x\.\d \bm{\Q}(\f,\f^\text{NL})\rt].
\ea
Since $\x^a$ is vanishing on the bifurcation surface $\mathcal{B}$, we have
\ba
\math{E}_{\S_1}(\f, \d \f^\text{NL})=- \int_\mathcal{B} {{\delta ^2}{\bm{Q}_\x }}=-\frac{\k_h}{8\p}\d^2A_\mathcal{B}^\text{NL}.
\ea
Therefore, the second-order inequality becomes
\ba\label{secorder1}
\d^2M-\F_\math{H}\d^2 Q\geq -\frac{\k_h}{8\p}\d^2A_\mathcal{B}^\text{NL}.
\ea
Here $A_\mathcal{B}^\text{NL}(\l)$ is the area of the bifurcation surface $\mathcal{B}$ for the static HDNL black hole with mass $M^\text{NL}(\l)$ and charge $Q^\text{NL}(\l)$.

For the left hand side of (\ref{secorder1}), according to (\ref{MQ}), (\ref{MQ2}) and (\ref{kAWF}), we have
\begin{equation}
\d^2M-\F_\math{H}\d^2 Q = \frac{{(n - 2)\Omega }}{{16\pi }} {{\delta ^2}m - \frac{{4\alpha q{{( - 2)}^{q - 1}}\Omega }}{{3 - n + \beta }}\frac{{{\sigma ^{\frac{1}{{2q - 1}}}}}}{{r_h^{3 - n + \beta }}}{{\delta ^2}\sigma}}
\end{equation}
where  $\sigma \equiv {C^{2q - 1}}$. From the line element \eq{ds2} of HDNL black holes, we  see that the right hand side of the inequality \eq{secorder1} can be calculated by taking two variations of the area formula $A_\mathcal{B}=\Omega r_h^{n-2}$. Using the fact that
\ba\label{exp22}
f(r_h^{{NL}}(\lambda ),m(\lambda ),\sigma(\lambda )) = 0
\ea
and taking the first-order variation of this equation, we  obtain
\ba\begin{aligned}
\delta r_h^{NL}\propto \d m-\frac{\d D}{r_h^{3-n+\b}},
\end{aligned}\ea
which implies that $\d r_h^\text{NL}=0$ under the optimal condition of the first-order perturbation inequality. By taking the second-order variation of equation \eq{exp22} and using the optimal condition $\d r_h^\text{NL}=0$,
we can further obtain
\ba\begin{aligned}
{\delta ^2}r_h^{{\rm{NL}}} = \frac{{r_h^{n + 1}{\delta ^2}{D^{{\rm{NL}}}}}}{{\beta r_h^nD - (n - 3)r_h^{\beta  + 3}m}}{\mkern 1mu}
\end{aligned}\ea
with
\ba\begin{aligned}
{D^{{NL}}}(\lambda ) = \frac{{\kappa \alpha {{( - 2)}^{q + 1}}{{(2{\rm{q}} - 1)}^2}{{(\sigma  + \lambda \sigma )}^{\frac{{2q}}{{2{\rm{q}} - 1}}}}}}{{(n - 2)(n - 2{\rm{q}} - 1)}}{\mkern 1mu},
\end{aligned}\ea
which implies that
\begin{equation}
{\delta ^2}{D^{{\rm{NL}}}} = {D^{{\rm{NL''}}}}(0) = \frac{{2\kappa \alpha q{{( - 2)}^{q + 1}}{\sigma ^{\frac{{2 - 2q}}{{2{\rm{q}} - 1}}}}}}{{(n - 2)(n - 2{\rm{q}} - 1)}}{\mkern 1mu} \delta {\sigma ^2}.
\end{equation}
Also, we have the the second-order variation of the area formula
\begin{align}
{\delta ^2}A_B^{{\rm{NL}}} &= (n - 2)\Omega r_h^{n - 3}( {(n - 3)r_h^{ - 1}{{(\delta r_h^{NL})}^2} + {\delta ^2}r_h^{NL}} ) \nonumber\\
&= (n - 2)\Omega r_h^{n - 3}{\delta ^2}r_h^{NL}.
\end{align}
By using the expression of the surface gravity
\ba\begin{aligned}
\kappa_h  = \frac{{(n - 3)m}}{{2r_h^{n - 2}}} - \frac{{\beta D}}{{2r_h^{\beta  + 1}}},
\end{aligned}\ea
and combining the above results, we can turn the right hand side of (\ref{secorder1})  into
\begin{equation}
 - \frac{\kappa_h }{{8\pi }}{\delta ^2}A_{\cal B}^{{\rm{NL}}} =\frac{{\kappa \alpha q{{( - 2)}^{q + 1}}\Omega {\sigma ^{\frac{{2 - 2q}}{{2{\rm{q}} - 1}}}}}}{{8\pi (n - 2{\rm{q}} - 1)r_h^{3 - n + \beta }}}{\mkern 1mu} \delta {\sigma ^2}.
\end{equation}

Finally, the second-order perturbation inequality becomes
\begin{equation}
{\delta ^2}m \ge \frac{{2\kappa \alpha q{{( - 2)}^{q + 1}}{\sigma ^{\frac{{2 - 2q}}{{2{\rm{q}} - 1}}}}}}{{(n - 2)(n - 2{\rm{q}} - 1)r_h^{3 - n + \beta }}}{\mkern 1mu} \left( {\delta {\sigma ^2} + (2q - 1)\sigma {\delta ^2}\sigma } \right).
\end{equation}
It can be further re-expressed as
\ba\begin{aligned}\label{secineq}
{\delta ^2}m - \frac{{{\delta ^2}D}}{{r_h^{3 - n + \beta }}} \ge 0.
\end{aligned}\ea
With the results of these two inequalities (\ref{ie11}) and (\ref{secineq}), we are now ready for the new version of Gedanken experiment.

\section{Gedanken experiments to destroy the nearly extremal black holes}\label{sec5}
Now we shall investigate the possibility in HDNL gravity to destroy the nearly extremal charged black holes by conducting the new version gedanken experiments.

Since we assume that the spacetime settles down to a static state in the asymptotic future, verifying the validity of the WCCC is equivalent to see whether there exists at least one root of the metric function or blackening factor $f(r(\l),m(\l),D(\l))$, which means that the line element still describes a black hole at sufficient late times. To make it computable, we define a function
\begin{equation}
h(\l)\equiv f(r_m(\l),m(\l),D(\l))=1-\frac{m(\l)}{r_m^{n-3}(\l)}+\frac{D(\l)}{r_m^\b(\l)}
\end{equation}
to describe the minimal value of the blackening factor in the asymptotic future. Here $r_m(\l)$ is the minimal radius of the blackening factor, and it can be obtained by
\ba\begin{aligned}\label{fprm}
{ \pd_rf(r_m(\l),m(\l),D(\l))=0}.
\end{aligned}\ea
{Using the explicit expression for the blackening factor, the above identity becomes
\ba\begin{aligned}\label{Mrm}
m(\l)=\frac{\b r_m^{n-\b-3}(\l) D(\l)}{(n-3)}.
\end{aligned}\ea
Under the zero-order approximation of $\l$, we have
\ba\begin{aligned}\label{Mqb}
m&=\frac{\b D r_m^{n-\b-3} }{(n-3)}.
\end{aligned}\ea
Taking the first-order variation to Eq. \eq{fprm}, we can further obtain
\ba\begin{aligned}
\d r_m&=\frac{r_m \d D}{\b D}.
\end{aligned}\ea}
Under the second-order approximation of perturbation, the minimal value of the blackening factor at late times can be expressed as
\begin{align}\label{frm}
h&(\l)\simeq 1-\frac{m}{r_m^{n-3}}+\frac{D}{r_m^\b}-\frac{\l}{r_m^{n-3}}\left(\d m-\frac{\d D}{r_m^{3+\b-n}}\right)-\nonumber\\
&\qquad-\frac{\l^2}{2r_m^{n-3}}\left(\d^2 m-\frac{\d^2 D}{r_m^{3+\b-n}}+\frac{(n-3)\d m\d r_m}{r_m}+\right.\nonumber\\
&\qquad\left.+\frac{\b\d r_m}{r_m^{\b+5-n}}[(3+\b-n)\d r_m-2 r_m\d D]\right)\,
\end{align}
{where we have used Eq.\eq{Mqb} to replace $m$ by $r_m, D$ and $\b$.}

 Since the gedanken experiments are only a perturbation for the background spacetime, it will only cause a small correction to the physical quantities at late time. Thus, in the aim to destroy the HDNL black hole, the initial state must be chosen as a nearly extremal black hole. We will consider the nearly extremal black hole situation for the background spacetime in the following. Then, the relation between the minimal value and horizon radius can be expressed as $r_m=(1-\epsilon)r_h$. With a similar setup as \cite{SW}, we assume that the parameter $\epsilon$ agree with the first-order approximation of perturbation.{ Then, we have
\begin{align}\label{frm2}
f(r_m)&=f((1-\e)r_h)\nonumber\\
&\simeq-\e r_h f'(r_h)+\frac{\e^2r_h^2}{2}f''(r_h)\nonumber\\
&\simeq-\e r_h f'(r_h)-\e^2 r_h^2 f''(r_h)+\frac{\e^2r_h^2}{2}f''(r_h)\nonumber\\
&=-\frac{\e^2r_h^2}{2}f''(r_h)\simeq -\frac{\e^2r_h^2}{2}f''(r_h)
\end{align}
under the second-order approximation of $\epsilon$, i.e., we have neglected the higher-order term $O(\e^3)$ of $\e$. We get
\begin{equation}
f_0=1-\frac{m}{r_m^{n-3}}+\frac{D}{r_m^\b}=\frac{\b(n-3-\b)D\e^2}{2r_h^\b}+O(\e^3)
\end{equation}
In the last step, we have replaced $r_m=(1-\e) r_h$ by $r_h$ and ignored the higher-order term $O(\e^3)$. For the first-order term in Eq. \eq{frm}, using the optimal condition of the first-order perturbation inequality, we have
\ba\begin{aligned}
f_1=\frac{(-n+3+\b)\l\d D \e}{r_h^\b}+O(\e^3,\l^3,\e^2\l,\cdots).
\end{aligned}\ea
For the second-order term, using the optimal condition of the first-order perturbation inequality and the second-order perturbation inequality, we can obtain
\ba\begin{aligned}
f_2=\frac{\l^2(n-3-\b)\d D^2}{2\b D r_h^{\b}}+O(\e^3,\l^3,\e^2\l,\cdots).
\end{aligned}\ea
Summing the above results, we reach to the final results
\ba\begin{aligned}
h(\l)\simeq -\frac{(3+\b-n)(\l\d D-\b D\e)^2}{2\b D r_h^\b}\leq 0
\end{aligned}\ea
this implies that the black hole cannot be destroyed under the second-order approximation. WCCC is also valid in the HDNL gravity scenario for the black hole solutions of the form \eq{ds2}\eq{ele}.

\section{Conclusion}\label{sec6}
We have studied the the validity of WCCC in HDNL black holes by considering the Sorce-Wald new version of gedanken experiment by the Iyer-Wald formalism.
 We have derived the first two perturbation inequalities in HNDL gravity, and then  conducted the new version of gedanken experiment. The result is that the nearly extremal static HNDL black holes cannot be overcharged under the second-order approximation. Therefore, there is no violation of the WCCC around the static black holes in HNDL gravity. This result  indicates that the validity of WCCC for more general higher dimensional nonlinear electrodynamic source and related systems.

\section*{acknowledgements}
The authors  thank Jie Jiang and Shupeng Song  for helpful discussions. ZL and XKG are partially supported by the National Natural Science Foundation of China through the
Grant Nos. 11875006 and 11961131013.
\\
\\

\end{document}